\documentclass[twocolumn]{jpsj2} 

\title{ Theory of Photoinduced Phase Transitions }

\author{
Kenji \textsc{Yonemitsu}$^{1,2}$\thanks{E-mail address: kxy@ims.ac.jp} and
Keiichiro \textsc{Nasu}$^{3,4}$
}

\inst{
$^1$Institute for Molecular Science, Okazaki, Aichi 444-8585 \\
$^2$Department of Functional Molecular Science, Graduate University for Advanced Studies, Okazaki, Aichi 444-8585 \\
$^3$Institute of Materials Structure Science, Oho 1-1, Tsukuba, Ibaraki 305-0801 \\
$^4$Department of Materials Structure Science, Graduate University for Advanced Studies, Oho 1-1, Tsukuba, Ibaraki 305-0801
}

\abst{
Theories of photoinduced phase transitions have developed along with the progress in experimental studies, especially concerning their nonlinear characters and transition dynamics. At an early stage, paths from photoinduced local structural distortions to global ones are explained in classical statistical models. Their dynamics are governed by transition probabilities and inevitably stochastic, but they were sufficient to describe coarse-grained time evolutions. Recently, however, a variety of dynamics including ultrafast ones are observed in different electronic states. They are explained in relevant electronic models. In particular, a coherent lattice oscillation and coherent motion of a macroscopic domain boundary need appropriate interactions among electrons and lattice displacements. Furthermore, some transitions proceed almost in one direction, which can be explained by considering relevant electronic processes. We describe the history of theories of photoinduced phase transitions and discuss a future perspective.
}

\kword{
photoinduced phase transition, molecular material, transition dynamics, coherent motion, neutral-ionic transition
}

\begin{document}
\maketitle

\section{Introduction} 

Nonequilibrium phases can be generated from an equilibrium one by external stimulations. Photoirradiation is one of them and sometimes allows the emergence of a new phase, which cannot be reached by simply changing temperature or pressure, because the energy of a photon is much higher than thermal energies. Although these phases finally relax to the starting equilibrium phase, they are locally stable, separated from the initial one by a substantial free-energy barrier (except for the transition from a Mott insulator to a metal, as will be mentioned later). When the relaxation time is long enough, the properties of this nonequilibrium phase can be studied by laser spectroscopy techniques, which have recently made great progress. Thus, photoinduced phase transitions have been studied extensively, both experimentally and theoretically. \cite{Nasu_book97,Nasu_book04,Nasu_rpp04} Here, we review progress in theories for such phase transitions. 

Photoirradiation creates electrons and holes, or excitons if they are bound. In most of the cases, they are accompanied by local structural deformation. To develop it spatially, some cooperativity is needed. One would ask first what kind of interaction possesses such cooperativity. Section 2 introduces classical models, in which photoexcited states proliferate in some conditions. Ground and excited states are simply assigned to Ising variables, so that the electronic transitions are statistically described on the basis of diabatic potentials. Section 3 first introduces adiabatic pictures for electronic models, which are useful in discussing necessary conditions for the spontaneous growth of photoexcited states. Transition dynamics do not always proceed adiabatically. Next, this section introduces the electronic dynamics derived from the time-dependent Schr\"odinger equation and shows their relevance to experimental findings. Section 4 concludes and presents future issues.

\section{Statistic Theory of Photoinduced Phase Transitions}

\subsection{Linear chain system}
\begin{figure}
\includegraphics[height=4cm]{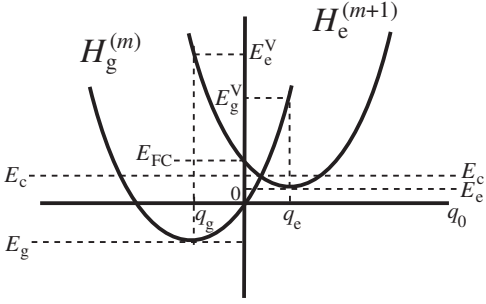}
\caption{Energy diagrams of 0th molecule, which is closest to cluster of $ m $ excited molecules, in ground and excited states. \cite{Hanamura_jpsj87,Nagaosa_prb89}}
\label{fig:two_parabolas}
\end{figure}

In order to clarify how the local structural distortions lead to global ones, a theory based on stochastic processes was proposed at first. Hanamura and Nagaosa introduced a simple model: \cite{Hanamura_jpsj87} 
\begin{align}
H =& \sum_l \mid e_l \rangle ( E_\mathrm{FC} - \sqrt{S} Q_l ) \langle e_l \mid 
  + \frac12 \sum_l Q_l^2 \nonumber\\ & - \frac12 \sum_{ll'} K_{ll'} Q_l Q_{l'}
\;, \label{eq:Hanamura}
\end{align}
and 
\begin{equation}
\mid g_l \rangle \langle g_l \mid + \mid e_l \rangle \langle e_l \mid = 1
\;,
\end{equation}
where $ E_\mathrm{FC} $ is the Franck-Condon excitation energy, $ Q_l $ the relevant displacement that dominantly interacts with an electron in the photoexcited $ \mid e_l \rangle $ state at the $ l $th molecule and is treated as a classical variable, and $ \sqrt{S} Q_l $ the Stokes shift in this state. The coupling strength between the displacements of the $ l $th and $ l' $th molecules is denoted by $ K_{ll'} $ with its diagonal element $ K_{ll} $ set at zero. 
Linear chain systems are first investigated in ref.~\citen{Hanamura_jpsj87}. Suppose that electrons on $ m $ consecutive molecules ({\it e.g.}, on sites 1 to $ m $) are in the excited state, and others are in the ground state. Each molecule is located on the respective equilibrium position. This state is governed by the Hamiltonian $ H_\mathrm{g}^{(m)} $. Then, consider that an electron on a neighboring molecule ({\it e.g.}, on site 0) is excited. The corresponding molecule is now away from the equilibrium position. This state is governed by the Hamiltonian $ H_\mathrm{e}^{(m+1)} $. The difference between these Hamiltonians is given by 
\begin{equation}
H_\mathrm{e}^{(m+1)} = H_\mathrm{g}^{(m)} + E_\mathrm{FC} - \sqrt{S} Q_0
\;.
\end{equation}
Then, an interaction mode $ q_0 $ (on site 0) is introduced, which is a linear combination of the displacements in the state with $ m $ excited molecules. The diabatic potentials for $ H_\mathrm{g}^{(m)} $ and $ H_\mathrm{e}^{(m+1)} $ are drawn as a function of $ q_0 $ (Fig.~\ref{fig:two_parabolas}).
They are displaced parabolas. The relative stability is determined by the minima of the two parabolas $ E_\mathrm{g} \equiv H_\mathrm{g}^{(m)}(q_0 = q_\mathrm{g}) $ and $ E_\mathrm{e} \equiv H_\mathrm{e}^{(m+1)}(q_0 = q_\mathrm{e}) $, where $ q_\mathrm{g} $ and $ q_\mathrm{e} $ give the respective minimum points, and by $ E^\mathrm{V}_\mathrm{g} \equiv H_\mathrm{g}^{(m)}(q_0 = q_\mathrm{e}) $ and $ E^\mathrm{V}_\mathrm{e} \equiv H_\mathrm{e}^{(m+1)}(q_0 = q_\mathrm{g}) $. If the inequality $ E_\mathrm{e} > E^\mathrm{V}_\mathrm{g} $ ($ E_\mathrm{g} > E^\mathrm{V}_\mathrm{e} $) is satisfied, the excited (ground) state at the 0th molecule is unstable and would decay into the ground (``excited'') state. Otherwise, the smaller one of $ E_\mathrm{g} $ and $ E_\mathrm{e} $ corresponds to the stable state, and the larger one to the metastable state. 
The analysis based on these relations reveals how a cluster of photoexcited molecules is stabilized by the electron-lattice interaction, whose dimensionless strength is the lattice relaxation energy $ S $ divided by $ E_\mathrm{FC} $, and by the coupling $ K $ among the displacements. A similar analysis also explains how a molecular excitation far from the cluster is attracted to the cluster. 

\subsection{Mean-field theory}
Nagaosa and Ogawa have extended this theory to general dimensions and considered the optical and thermal transitions. \cite{Nagaosa_prb89} When the relaxation of the interaction mode $ Q_l $ is rapid, the system is always in equilibrium with a given electronic state. These modes can be integrated out to reduce the original model (\ref{eq:Hanamura}) to the Ising model, where the spin-up and spin-down states correspond to the excited and ground states: 
\begin{equation}
H = -h \sum_l \sigma_l - \frac12 \sum_{l\neq l'} J_{ll'} \sigma_l \sigma_{l'}
\;,
\end{equation}
where the field $ h $, the exchange $ J_{ll'} $, and the phonon dispersion $ \omega_k $ are given by 
\begin{equation}
h = - E_\mathrm{FC} + \frac{S}{2(1-K)}
\;,
\end{equation}
\begin{equation}
J_{ll'} = \frac{S}{N} \sum_{k} 
\frac{\cos[\mathbf{k}\cdot(\mathbf{R}_l-\mathbf{R}_{l'})]}{\omega_k^2}
\;,
\end{equation}
\begin{equation}
\omega_k^2 = 1 - \sum_{l'} K_{ll'} 
\exp[\mathrm{i} \mathbf{k}\cdot(\mathbf{R}_l-\mathbf{R}_{l'})]
\;,
\end{equation}
with $ \mathbf{R}_l $ being the lattice vector of the $ l $th molecule, and $ \omega_{k=0}^2 = 1 - K $. 
The intersite couplings are treated in the mean-field approximation. For instance, the thermal transition probabilities were assumed to be the product of the attempt frequency and the activation factor, 
$ P^\mathrm{th}_{{\rm e}\rightarrow{\rm g}} = 
\tau^{-1} \exp\left[-(E_\mathrm{c}-E_\mathrm{e})/(k_\mathrm{B}T)\right] $ and 
$ P^\mathrm{th}_{{\rm g}\rightarrow{\rm e}} = 
\tau^{-1} \exp\left[-(E_\mathrm{c}-E_\mathrm{g})/(k_\mathrm{B}T)\right] $, 
where $ E_\mathrm{c} $ is the energy of the intersection of the two parabolas. The stochastic dynamics of the system under and after the optical pumping are investigated on the basis of the rate equation for the density of the excited states $ n_\mathrm{e}(t) $, which is a function of time $ t $. When stable and metastable states are present, it is written as 
\begin{equation}
\frac{\mathrm{d}}{\mathrm{d}t} n_\mathrm{e}(t) = I(t) 
+ P^\mathrm{th}_{{\rm g}\rightarrow{\rm e}} [ 1 - n_\mathrm{e}(t) ]
- P^\mathrm{th}_{{\rm e}\rightarrow{\rm g}} n_\mathrm{e}(t)
\;,
\end{equation}
where $ I(t) $ is the optical pumping term, being positive to create the $ e $ state or negative to create the $ g $ state. The activation energies in the exponents of the thermal transition probabilities are quadratic functions of $ n_\mathrm{e}(t) $ in the mean-field approximation. 
The time evolution of $ n_\mathrm{e}(t) $ after $ n_\mathrm{e}(0) $=0 depends on the strength $ I_0 $ and duration $ t_0 $ of the optical pumping. When $ I_0 $ is less than a threshold value $ I_\mathrm{th} $, however long the optical pumping is applied, $ n_\mathrm{e}(t) $ does not exceed a certain value. After the pumping is switched off, the system always relaxes to the initial state $ n_\mathrm{e}(0) $=0. When $ I_0 $ is greater than $ I_\mathrm{th} $, on the other hand, $ n_\mathrm{e}(t) $ monotonically increases without saturation for $ 0 < t < t_0 $. After the pumping is switched off, $ n_\mathrm{e}(t) $ relaxes to the final value $ n_f $=0 for $ t_0 < t_\mathrm{cr} $ and to $ n_f $=1 for $ t_0 > t_\mathrm{cr} $. The critical value of the duration time $ t_\mathrm{cr} $, above which the system is switched from $ n_i $=0 to $ n_f $=1, is inversely proportional to $ I_0 $ for large $ I_0 $ and diverges as $ t_\mathrm{cr} \sim (I_0-I_\mathrm{th})^{-1/2} $ near the diverging point $ I_\mathrm{th} $. The relaxation times are also discussed. The theory is applied to photoinduced structural transformations in polydiacetylenes.

\subsection{Other theories}
The theories above are applied to systems with weakly interacting components to which thermal fluctuations are so dominant that the transition probability is a key quantity. Spin-crossover complexes are such examples, where spins indirectly interact with each other through couplings with lattice displacements. The Hanamura-Nagaosa theory is extended to systems with internal degrees of freedom, {\it e.g.}, spin-crossover complexes that are composed of dimers and show two-step transitions. \cite{Luty_jpsj04} 
In general, stochastic dynamics can be treated in classical models with master equations \cite{Romstedt_jpcs98,Boukheddaden_epjb00,Hoo_epjb00,Nishino_prb01} and Monte Carlo simulations. \cite{Romstedt_jpcs98,Nishino_prb01,Kawamoto_apl02,Sakai_jpsj02}

\section{Electronic Theories of Photoinduced Phase Transitions}

\subsection{Domino effect}
\begin{figure}
\includegraphics[height=1.5cm]{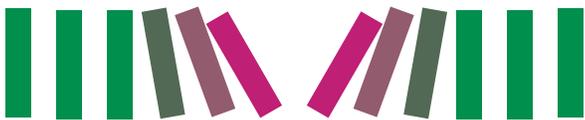}
\caption{(Color online) Schematic structural transformation. The standing and falling dominoes correspond to the displacements accompanied with stable and metastable localized electronic states.}
\label{fig:domino}
\end{figure}

Photoinduced dynamics are sometimes regarded as domino effects. This description is inappropriate for systems showing coherent dynamics, but gives a hint for the mechanisms of photoinduced phase transitions. Koshino and Ogawa theoretically show a domino-like structural transformation in a one-dimensional model composed of localized electrons and classical lattice displacements: \cite{Koshino_jpsj98} 
\begin{align}
H = & \sum_j \left[ 
\left(
\begin{array}{cc}
\epsilon - 2 \gamma u_j & t_0 \\ t_0 & 2 u_j
\end{array}
\right)
+ u_j^2 + \left( \frac{\partial u_j}{\partial t} \right)^2 
\right] \nonumber\\
& + \sum_{(i,j)} k_{ij} ( u_i - u_j )^2
\;,
\end{align}
where $ u_j $ and $ \partial u_j/\partial t $ denote the $ j $th displacement and velocity, respectively, $ k_{ij} $ the elastic constants,  $ \epsilon $ the energy difference between the two electronic states at $ u_j $=0, $ t_0 $ the overlap integral between them, and $ \gamma $ the electron-lattice coupling constant. 
After a photon is absorbed by an electron at a site, the lattice locally relaxes under friction to the minimum on the adiabatic potential. Then, after a photon is spontaneously emitted, three evolution patterns are shown to exist. When the elastic couplings $ k_{ij} $ are too weak, the local structural distortion remains locally. When they are too strong, the initial local distortion is pulled back by surrounding lattice displacements to the position before the absorption. Only when they are intermediately strong and short-ranged will the initial local distortion trigger a global structural transformation as the domino effect (Fig.~\ref{fig:domino}). 

\subsection{Adiabatic relaxation path}
Photoinduced phase transitions are often related to multistability, where different electronic phases are stable and metastable. Such stability may be manifested by the presence of a first-order phase transition induced by changing temperature or pressure. The metastable phase may be hidden in equilibrium so that it is manifested only by photoirradiation, which gives a much higher energy to the system than thermal energies. The photoinduced phase transition dynamics depend on the electronic state and especially on how the initial state is prepared. For this purpose, relevant itinerant-electron models are necessary. They have finite transfer integrals between neighboring molecular orbitals. They are off-diagonal elements of the model Hamiltonian, giving transition {\it amplitudes}. In contrast, the stochastic dynamics within classical statistical models are governed by transition {\it probabilities}, which are often given by Boltzmann factors satisfying detailed balance. 

Organic molecular tetrathiafulvalene-chloranil (TTF-CA) crystals are the most intensively studied. The donor TTF and acceptor CA molecules are alternately stacked along the most conducting axis. At low temperature or under high pressure with contraction, these molecules become ionic due to the long-range Coulomb interaction. \cite{Torrance_prl81} Otherwise, they are neutral due to the large difference between the ionization potential of the donor molecule and the electron affinity of the acceptor molecule. In the ionic phase at ambient pressure, these molecules are dimerized. Both ionic-to-neutral and neutral-to-ionic transitions are photoinduced, as shown by the optical reflectivity. \cite{Koshihara_jpcb99} It is found that photons with 2.3 eV corresponding to intramolecular excitations at TTF sites are much more efficient for generation of neutral domains in the ionic background than those with 0.6 eV corresponding to intrachain charge-transfer excitations. \cite{Suzuki_prb99} The latter clearly have a threshold in the photoexcitation density, below which a macroscopic neutral domain cannot be generated. Thus, the photoinduced dynamics are very sensitive to the initial condition of the electronic state, from which the lattice relaxation starts. 

These observations have motivated theoretical studies based on itinerant-electron models. Huai {\it et al.} have clarified the adiabatic relaxation path from the Franck-Condon-type photoexcited state above the ionic ground state to the formation of a large neutral domain, using a one-dimensional extended Peierls-Hubbard model with alternating potentials (and the interchain interaction mentioned later): \cite{Huai_jpsj00} 
\begin{align}
H = & -t_0 \sum_{l,\sigma}
   \left( c^{\dagger}_{l,\sigma}c_{l+1,\sigma} + \mathrm{h.c.} \right)
 + \frac{\Delta}{2} \sum_{l} (-1)^{l} n_{l} \nonumber \\
& + U \sum_{l} n_{l,\uparrow} n_{l,\downarrow} 
+ \sum _{l} V_{l}(q_l,q_{l+1}) \delta n_{l} \delta n_{l+1} \nonumber \\
& + \sum _{l} \frac{S_1}{2}( q_{l} - q_{l+1} )^{2}
+ \sum _{l} \frac{S_2}{4}( q_{l} - q_{l+1} )^{4}
\;, \label{eq:Huai}
\end{align}
where $ c^{\dagger}_{l,\sigma} $ ($ c_{l,\sigma} $) is the creation (annihilation) operator of an electron with spin $\sigma$ at site $l$, $ n_{l,\sigma} = c^{\dagger}_{l,\sigma} c_{l,\sigma} $, $ n_{l} = n_{l,\uparrow} + n_{l,\downarrow} $, $ \delta n_{l} = n_{l} - 2 $ for odd $l$, $ \delta n_{l} = n_{l} $ for even $l$, $ q_{l} $ is the dimensionless displacement of the $l$th molecule along the chain from its equidistant position. The distance between the $l$th and $(l+1)$th molecules is given by $ d_{l,l+1} = d_0 (1+q_{l+1}-q_l) $, where $d_0$ is the average intermolecular distance. The donor and acceptor molecules are located at odd and even sites, respectively. The nearest-neighbor repulsion strength is assumed to depend nonlinearly on the intermolecular distance as 
\begin{equation}
V_{l}(q_l,q_{l+1}) = V_0 + 
\beta_{1} ( q_{l} - q_{l+1} ) + \beta_{2} ( q_{l} - q_{l+1} )^{2}
\;,
\end{equation}
where $ V_0 $ is for the regular lattice, and $ \beta_{1} $ ($ \beta_{2} $) is the linear (quadratic) coefficient. 
The parameter $ t_0 $ denotes the transfer integral, $ \Delta $ the level difference between the neighboring orbitals in the neutral limit, and $ U $ the on-site repulsion strength. The elastic energy is expanded up to the fourth order: the parameters $ S_{1} $ and $ S_{2} $ are the linear and nonlinear elastic constants. 

\begin{figure}
\includegraphics[height=4cm]{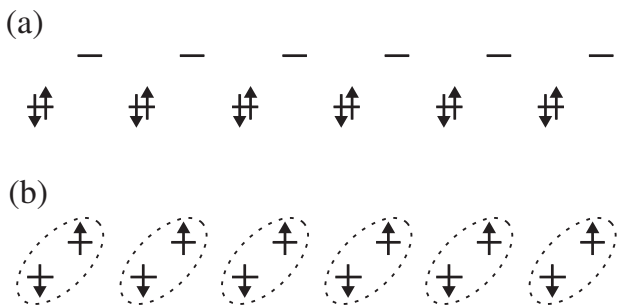}
\caption{Schematic electronic and lattice structures in (a) neutral and (b) ionic phases.}
\label{fig:NI}
\end{figure}

\begin{figure}
\includegraphics[height=3.5cm]{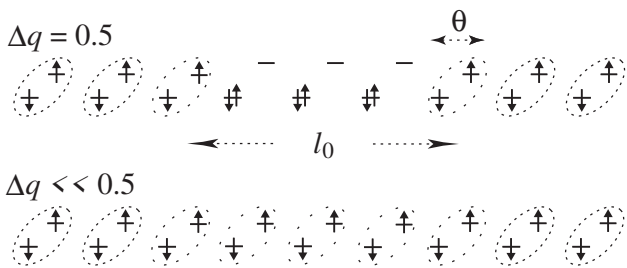}
\caption{Schematic excited domains in ionic background.}
\label{fig:N_in_I}
\end{figure}

In the neutral phase, the orbital of the donor molecule is almost doubly occupied, while that of the acceptor molecule is almost empty [Fig.~\ref{fig:NI}(a)]. The total charge of the donor molecule at site $ l $ is $ +(2- n_l)= - \delta n_{l} $, while that of the acceptor molecule is $ -n_l = - \delta n_{l} $. In the ionic phase, both orbitals are almost singly occupied [Fig.~\ref{fig:NI}(b)]. 
The values of the eight parameters are so determined that they reproduce the {\it ab initio} estimation of the transfer integral, the ionicity in the ionic phase, that in the neutral phase, the degree of dimerization in the ionic phase, the energies and the relative strength of the charge-transfer absorption peaks in the ionic phase, and the charge-transfer absorption energy in the neutral phase. The lattice relaxation path starting from the Franck-Condon state with a single charge-transfer exciton is assumed to be described by a domain, 
\begin{equation}
q_l = (-1)^l q_0 \left\{ 
1 + \Delta q \left[
\tanh \left( 
\theta ( \mid l \mid - \frac{l_0}{2} )
\right) - 1 \right]\right\}
\;,
\end{equation}
where $ (-1)^l q_0 $ denotes the uniform dimerization in the ionic state, $ \Delta q $ the amplitude of a local distortion induced by an excited domain, $ \theta $ the width of the domain boundary, and $ l_0 $ the domain size (Fig.~\ref{fig:N_in_I}). 
An interchain interaction between displacements is introduced to confine the domain in the ionic background: 
\begin{equation}
H_\mathrm{inter}^\mathrm{ph} = \sum_l \sum_{i=1,2,3} 
K_i \left( q_l - (-1)^l q_0 \right)^{2i} 
\;,
\end{equation}
where $ q_l $ denotes the displacement of a central chain surrounded by ionic environmental chains, and $ K_i $ the 2$ i $th coefficient with respect to the distortion in the central chain. 

The adiabatic energy surface of the ground state and that of the first excited state are drawn as a function of the amplitude $ \Delta q $ and the size $ l_0 $ of the domain with the width of the domain boundary $ \theta $ optimized. \cite{Huai_jpsj00} The energy surface of the ground state has a plateau corresponding to a neutral domain separated from the ground state by a low potential barrier. In the energy surface of the first excited state, a local minimum corresponding to the charge-transfer exciton appears around the Franck-Condon state. Another local minimum appears and corresponds to a neutral domain, which are separated from the Franck-Condon state by such a high potential barrier that the absorption of a single photon with 0.6 eV cannot overcome it. The domain is stable only when its size exceeds a critical value because the energy required for creation of the two domain boundaries makes a small domain unstable in the excited state. The spin- and charge-density distributions of the ground and excited states with a neutral domain are very similar: they are different only at the domain boundaries. Because a single charge-transfer exciton cannot trigger the formation of a macroscopic neutral domain, a large excess energy is needed at the very beginning of the relaxation to overcome the high barrier, for charge-transfer excitons to proliferate, and finally to form a macroscopic neutral domain. The process described above is only for the early nucleation stage of the photoinduced macroscopic transformation. 

\subsection{Transition dynamics in TTF-CA}
Along with the progress in experimental techniques, the pump-probe reflection spectroscopy acquires higher time resolution. Ultrafast optical switching from the ionic to the neutral state is observed. \cite{Iwai_prl02} Neutral strings are produced within 2 ps after a resonant excitation of the charge transfer band at 0.65 eV polarized along the stacking axis. Just below the transition temperature, a macroscopic conversion from the ionic to the neutral states is achieved within 20 ps. It is accompanied with coherent motion of the macroscopic neutral-ionic domain boundary with a much longer period than that of the optical lattice oscillation. 

\begin{figure}
\includegraphics[height=5cm]{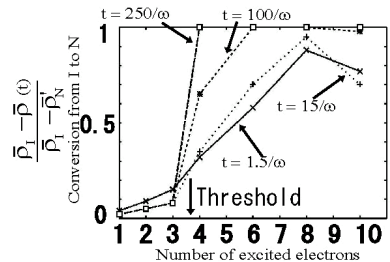}
\caption{Conversion from ionic to neutral phases, as a function of number of excited electrons in 100-site chain. \cite{Miyashita_jpsj03} The nonlinearity is strengthened with the passing of time after the photoexcitation, $ t $=1.5$/\omega$, 15$/\omega$, 100$/\omega$ and 250$/\omega$, where $ \omega $ denotes the bare phonon frequency. The initial lattice temperature is $ T $=$ 10^{-2}t_0 $.}
\label{fig:I_to_N_fraction}
\end{figure}

Theoretically, real-time dynamics need to be calculated for further understanding. First, results are obtained by solving the time-dependent Schr\"odinger equation of an electron-lattice model without electron-electron interaction, where two charge-density-wave states with opposite polarizations are degenerate. \cite{Iwano_prb02} The lattice part is treated classically. The photoconversion is shown to be nonlinear. During the photoconversion process, macroscopic oscillations are observed, which are essentially different from the linear modes in equilibrium. Then, by adding the kinetic energy of the displacements, 
\begin{equation}
H_\mathrm{kin} = \frac12 m_{l} \dot{q}_{l}^{2}
\;,
\label{eq:kin}
\end{equation}
with the $ l $th molecular mass $ m_l $, to the model (\ref{eq:Huai}), Miyashita {\it et al.} have calculated real-time dynamics of ionicity and dimerization within the unrestricted Hartree-Fock approximation after the occupancy of orbitals in the ionic ground state is initially changed. \cite{Miyashita_jpsj03} As the number of orbitals whose occupancy is changed increases, i.e., with increasing photoexcitation density, more neutral domains are created. Above the threshold excitation density, the neutral phase is finally achieved. After the photoexcitation, ionic domains with wrong polarization generally appear, which reduce the correlation length of the staggered lattice displacement. As the degree of initial lattice disorder increases, more solitons appear between the ionic domains with opposite polarizations, which obstruct the growth of neutral domains and slow down the transition. From the Fourier transform of the ionicity as a function of time, three characteristic time scales are observed: rapid charge-transfer excitations corresponding to the peak energy of a charge-transfer exciton in the random-phase-approximation spectra, slow charge-transfer excitations strongly coupled with optical lattice oscillations, and much slower excitations due to the collective motion of neutral-ionic domain boundaries. The last corresponds to the experimentally observed, coherent motion of the macroscopic neutral-ionic domain boundary, \cite{Iwai_prl02} and is shown to be sensitively affected by collisions with the solitons. When the fraction of conversion from the ionic to the neutral phases is plotted as a function of the number of excited electrons, the curve evolves with time (Fig.~\ref{fig:I_to_N_fraction}). 
Immediately after the photoexcitation, it is close to a linear function. Above the threshold number, the fraction monotonically increases with time until the photoinduced phase transition is completed. Below the threshold, the fraction decreases and eventually becomes very small. This result is consistent with the experimental finding. \cite{Iwai_prl02} 

More recently, experimental results have been summarized covering both the ionic-to-neutral and the neutral-to-ionic transitions with different excitation energies, excitation densities and temperatures. \cite{Okamoto_prb04} The ionic-to-neutral transition induced by the resonant excitation of the charge-transfer band proceeds with (1) initial formation of a confined one-dimensional neutral domain, (2) multiplication of such domains to semi-macroscopic neutral states by 20 ps, and (3) evolution in the direction normal to the sample surface. The dynamics of the neutral-to-ionic transition that is induced by a similar excitation are clearly different from the above. Although one-dimensional ionic domains are initially produced by lights, they decay within 20 ps even if the excitation density is high. The initial conversion fraction is a linear function of the excitation density. 

Thus, we are motivated to incorporate a pulse of oscillating classical electric field,
\begin{equation}
E(t) = E_\mathrm{ext} \sin \omega_\mathrm{ext} t
\;,
\end{equation}
with amplitude $ E_\mathrm{ext} $ and frequency $ \omega_\mathrm{ext} $ for $ 0 < t < 2 \pi N_\mathrm{ext} / \omega_\mathrm{ext} $ with integer $ N_\mathrm{ext} $, into the Peierls phase of the transfer integral. The time-dependent Schr\"odinger equation is solved again for the one-dimensional extended Peierls-Hubbard model with alternating potentials. Now, the frequency, the amplitude and the duration of the pulse can be explicitly and independently varied. The dynamics of the ionic-to-neutral transition are indeed qualitatively different from those of the neutral-to-ionic transition. 

\subsection{Difference between I-to-N and N-to-I transitions}
\begin{figure}
\includegraphics[height=5cm]{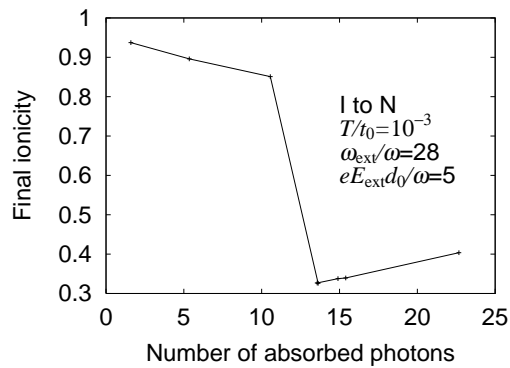}
\caption{Final ionicity as a function of number of absorbed photons in 100-site chain, when electric field of strength $ eE_\mathrm{ext}d_0 $=5$ \omega $ and of frequency $ \omega_\mathrm{ext} $=28$ \omega $ below linear-absorption peak at 32$ \omega $ is applied to {\it ionic} phase. \cite{Yonemitsu_jpsj04a} $ \omega $ denotes the bare phonon frequency. The initial lattice temperature is $ T $=$ 10^{-3}t_0 $.}
\label{fig:I_to_N_threshold}
\end{figure}

When the dimerized ionic phase is photoexcited, the threshold behavior is observed again by plotting the final ionicity as a function of the increment of the total energy, i.e., as a function of the number of absorbed photons (Fig.~\ref{fig:I_to_N_threshold}). \cite{Yonemitsu_jpsj04a} 
Above the threshold photoexcitation density, the system finally enters a neutral state with equidistant molecules. This is the case even after the electric field is switched off. Once produced by the field, a neutral domain grows spontaneously. Its growth cannot be stopped by switching off the field. When compared with the threshold behavior derived from the rate equation, \cite{Nagaosa_prb89} we do not observe a quantity corresponding to the threshold intensity $ I_\mathrm{th} $. Even if the amplitude of the pulse is small (i.e., even if the pulse is weak), a transition seems finally achieved by setting the pulse duration long enough. The deterministic dynamics are more efficient than the stochastic ones because of the restricted energy dissipation. When compared with the domino effect, the motion of the neutral domain is similar in that its growth is spontaneous once triggered. However, it is quite different from the motion of dominoes under strong friction on the adiabatic potential, \cite{Koshino_jpsj98} which is far from being coherent. 

For a given amount of the increment of the total energy, the lattice dynamics relative to the charge dynamics in the ionic-to-neutral transition depend on the character of the pulse. When the pulse is strong and short, the charge transfer takes place on the same time scale with the disappearance of dimerization. When the pulse is weak and long, the dimerization-induced ferroelectric polarization is disordered first to restore the inversion symmetry, and then the charge transfer takes place to bring the system to a neutral state. The difference between the two time scales increases as the pulse is weakened. Between these time scales, a paraelectric ionic phase is realized. In the case of intramolecular excitations at 1.55 eV, such a new ionic phase with disordered polarizations is suggested to appear by time-resolved X-ray diffraction. \cite{Guerin_cp04} This similarity may be due to the fact that, when the pulse is weak, the supplied energy is not directly used to transfer charge density along the chain. 
It is furthermore shown theoretically that infrared light polarized along the chain can also induce the ionic-to-neutral transition if the excitation density exceeds a critical value, which is currently beyond the value experimentally achieved. 

\begin{figure}
\includegraphics[height=5cm]{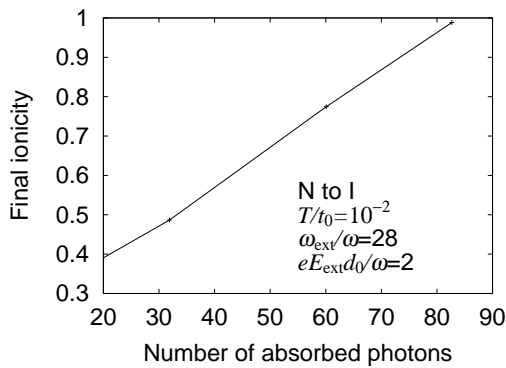}
\caption{Final ionicity as a function of number of absorbed photons in 100-site chain, when electric field of strength $ eE_\mathrm{ext}d_0 $=2$ \omega $ and of frequency $ \omega_\mathrm{ext} $=28$ \omega $ below linear-absorption peak at 30$ \omega $ is applied to {\it neutral} phase. \cite{Yonemitsu_jpsj04b} $ \omega $ denotes the bare phonon frequency. The initial lattice temperature is $ T $=$ 10^{-2}t_0 $.}
\label{fig:N_to_I_linear}
\end{figure}

When the neutral phase is photoexcited, the linear behavior is observed by plotting the final ionicity as a function of the increment of the total energy (Fig.~\ref{fig:N_to_I_linear}). \cite{Yonemitsu_jpsj04b} 
The neutral-to-ionic transition proceeds in an uncooperative manner. If the oscillating electric field is turned off before the transition is completed, the ionicity remains intermediate unless the energy dissipation is taken into account. The growth of a metastable domain is not spontaneous but forced by the external field. The final state is determined merely by how many photons are absorbed to increase the ionicity. This result is consistent with the experimental finding in the neutral-to-ionic transition induced by intrachain charge-transfer photoexcitations. \cite{Okamoto_prb04} By intramolecular excitations, on the other hand, a transition from the neutral phase to the ionic phase with three-dimensionally ordered ferroelectric polarizations is shown to be induced, as clearly demonstrated by X-ray diffraction. \cite{Collet_science03} The reason for this difference is unclear yet. 

The qualitative difference described above between the two transitions is not due to the different inversion symmetry between the dimerized ionic and the regular neutral states because the difference survives even if the electron-lattice coupling is turned off so that both ionic and neutral states have even parity with respect to the space inversion. The most plausible reason for the difference is the fact that the ionic state is a Mott insulator caused by the electron correlation, whereas the neutral state is a band insulator caused by the band structure. In a Mott insulator, all the electrons are so correlated that any one of them cannot easily make the first move. However, above the threshold excitation density, they cannot tolerate the increased total energy. Once some electrons move, they trigger the collective motion. One electron is transferred after another like dominoes. The collective dynamics in the ionic phase are due to the electron-electron interaction that realizes the Mott insulator phase. They cannot be described by stochastic approaches of refs.~\citen{Hanamura_jpsj87} and \citen{Nagaosa_prb89}. In a band insulator, electrons move individually. The supplied energy is merely consumed to transfer electrons from the donor to acceptor molecules almost independently. They do not strongly influence the motion of other electrons. 

Thus far, we have focused on the dynamics in purely one-dimensional systems. Although the short-time behavior may not suffer from interchain interactions, they are important for the condition of coherence, which can be manifested by a double pulse. \cite{Yonemitsu_jpsj04c} The coherent motion of the macroscopic neutral-ionic domain boundary \cite{Iwai_prl02} is possible only when interchain interactions are strong enough. Although the effect of interchain elastic couplings is considered in the context of confinement of a metastable domain, \cite{Huai_jpsj00} interchain electron-electron interactions are much stronger. The intramolecular charge distribution in the TTF and CA molecules is calculated by the {\it ab initio} quantum chemical method. \cite{Kawamoto_prb01} The interchain electrostatic energies between neighboring molecules are smaller than but comparable to the intrachain ones. Because the molecules are tilted, the relative repulsion strengths are very different from the naive expectations based on the intermolecular distance measured in terms of the center of mass of each molecule. In the ionic phase, the interchain attractive interaction between the neighboring donor and acceptor molecules along the $ b' $ axis is larger than any interchain repulsive one between the donors or between the acceptors. The consequent electrostriction is theoretically shown \cite{Kishine_prb04} to be responsible for the pressure-temperature phase diagram of the TTF-CA complex containing the paraelectric ionic phase \cite{Lemee_prl97} and for the discontinuous contraction along the $ b $ axis. \cite{Luty02} 

Then, we add to the model (\ref{eq:Huai}) with (\ref{eq:kin}) interchain electron-electron interactions, \cite{Yonemitsu_jp_jltp_unpublished}
\begin{equation}
H_\mathrm{inter}^\mathrm{el} = \sum_{l,j} \left(
   U_{\perp}  \delta n_{l,j} \delta n_{l,j+1} 
 + V_{\perp1} \delta n_{l,j+1} \delta n_{l+1,j} 
\right)
\;,
\end{equation}
where $ j $ is the chain index, and the donor-acceptor coupling strength $ V_{\perp1} $ is slightly larger than the donor-donor or acceptor-acceptor one $ U_{\perp} $. \cite{Kawamoto_prb01} 
When the ionic phase is photoexcited, the transition dynamics depend on the strengths of the interchain couplings. With weak interchain couplings, for instance, halves of the values for TTF-CA, the interchain correlation is very weak during the transition. A first neutral domain is easily created by a low density of photons. If the electric field is switched off immediately after this absorption, the residual chains remain ionic. By continuing the application of the electric field, next domains are created in the neighboring chains. Many photons are finally needed to complete the transition into the neutral phase. With strong interchain couplings comparable to those in TTF-CA, the interchain correlation is strong during the transition. To create a first neutral domain, more photons need to be absorbed. Once a domain is nucleated, nearby domains are almost simultaneously created in the neighboring chains to grow cooperatively. Their growth cannot be stopped even if the electric field is switched off immediately after the appearance of the first domain. As a consequence, the density of absorbed photons needed to complete the transition is lower than in the weak-coupling case. Because of the strong interchain correlation, the growth of metastable domains coherently proceeds, which is consistent with the experimentally observed, coherent motion of the macroscopic neutral-ionic domain boundary. \cite{Iwai_prl02}

\subsection{Photoinduced phases in MX and MMX chains}
Besides the TTF-CA complex described above, many other molecular materials have been or are currently studied to show their own transition dynamics. Some of them show very asymmetric transitions between two different electronic phases. 
In a halogen-bridged nickel-chain compound, the ultrafast photoinduced Mott transition from a charge-transfer insulator to a metal is observed. \cite{Iwai_prl03} Photoinduced optical responses of one-dimensional half-filled Hubbard models with and without alternating potentials are studied by the exact diagonalization method to show a large shift of the spectral weight to low energies including the Drude component in the Mott insulator phase, which are similar to the response caused by chemical doping. \cite{Maeshima_jpsj05} 

\begin{figure}
\includegraphics[height=4.4cm]{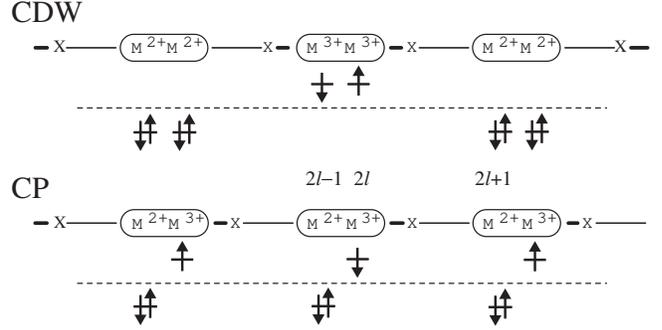}
\caption{Schematic electronic and lattice structures in MMX chains.}
\label{fig:CDW_CP}
\end{figure}

In halogen-bridged binuclear platinum-chain compounds, R$_4$[Pt$_2$(pop)$_4$I]$ n $H$_2$O (pop=P$_2$O$_5$H$_2^{2-}$) with cation R, the transition from a charge density wave (CDW) to a charge-polarization (CP) state is photoinduced much more easily than its inverse transition. \cite{Matsuzaki_prl03} The electronic phases of these compounds are controlled by the distance between the neighboring binuclear units $ d_\mathrm{MXM} $ both when the counterion R is varied and when the applied pressure is varied. The nonmagnetic CDW phase is observed for small $ d_\mathrm{MXM} $, while the paramagnetic CP phase for large $ d_\mathrm{MXM} $.
The charge density is disproportioned among binuclear units in the CDW phase, and within each binuclear unit in the CP phase (Fig.~\ref{fig:CDW_CP}). In both phases, the iodine ion approaches the platinum ion with less electron density. 
The dependences of the electronic phase and its physical properties on $ d_\mathrm{MXM} $ are well explained by perturbation theories from the strong-coupling limit as well as the exact diagonalization method. \cite{Kuwabara_jmc01}
The mechanisms of the photoinduced phase transitions are clarified by taking the charge transfer processes explicitly into account \cite{Yonemitsu_prb03} in a one-dimensional two-band three-quarter-filled Peierls-Hubbard model, 
\begin{align}
H = & - t_{\rm MM} \sum_{l, \sigma} ( c^\dagger_{2l-1, \sigma} c_{2l, \sigma} + \mathrm{h.c.} ) \nonumber \\ & 
- t_{\rm MXM} \sum_{l, \sigma} ( c^\dagger_{2l, \sigma} c_{2l+1, \sigma} + \mathrm{h.c.} ) \nonumber \\ & 
 + \beta \sum_l q_l ( n_{2l+1} - n_{2l} ) 
 + U_{\rm M} \sum_l n_{l, \uparrow} n_{l, \downarrow} \nonumber \\ & 
 + K_{\rm MX} \sum_l q_l^2 + (M/2) \sum_l \dot{q}_l^2 \;,
\end{align}
where $ c^\dagger_{l, \sigma} $ creates an electron with spin $ \sigma $ at site $ l $, $ n_{l, \sigma} = c^\dagger_{l, \sigma} c_{l, \sigma} $, and $ n_l = \sum_\sigma n_{l, \sigma} $. The binuclear unit contains two M (M=Pt) sites, 2$ l-$1 and 2$ l $. The displacement of the X (X=I) ion between the two M sites 2$ l $ and 2$ l $+1, relative to that in the undistorted structure, is denoted by $ q_l $. The transfer integral within the unit is denoted by $ t_{\rm MM} $, while that between the neighboring units through the X $ p_z $ orbital by $ t_{\rm MXM} $. The energy level of the M $ d_{z^2} $ orbital depends on $ q $ with linear coefficient $ \beta $. The on-site repulsion strength is denoted by $ U_{\rm M} $, the elastic constant for the MX bond by $ K_{\rm MX} $, and the mass of the X ion by $ M $. With increasing pressure corresponding to decreasing $ d_\mathrm{MXM} $, $ \beta $ increases because it originates from the electrostatic energy between the positive M and the negative X ions. This is why the discontinuous transition from the CP to the CDW phases is induced by pressure. \cite{Matsuzaki_prl03}

\begin{figure}
\includegraphics[height=5.5cm]{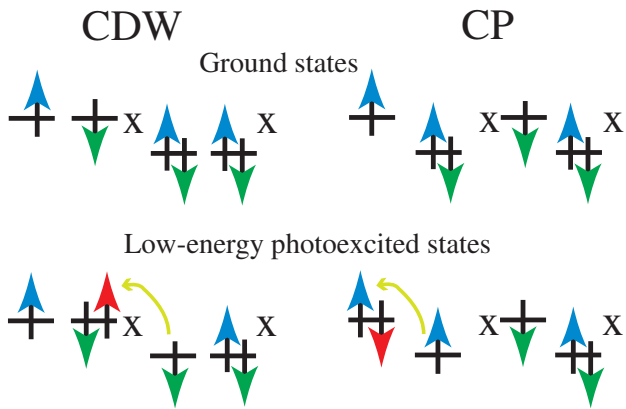}
\caption{(Color online) Schematic ground and low-energy photoexcited states in MMX chains. \cite{Yonemitsu_prb03} In the CDW phase, the charge is transferred to the neighboring unit to form a transient CP domain. In the CP phase, the charge is transferred within the unit to merely reverse the polarization.}
\label{fig:CDW_to_CP}
\end{figure}

It is true that the dependence of the threshold excitation density in the photoinduced CDW-to-CP transition on the pressure, which controls the relative stability of these phases, can be explained qualitatively with their diabatic potentials, i.e., in a stochastic manner. However, the asymmetry in these transitions requires explanation based on the explicit charge transfer processes (Fig.~\ref{fig:CDW_to_CP}): in the CP state, low-energy charge-transfer processes occur only within a binuclear unit, and consequently they do not lead to the CDW state. 
Even if the CP phase is irradiated by higher-energy photons, charge transfer among the binuclear units takes place incoherently. Since the interunit charge transfer is energetically unfavorable, the CDW domains never proliferate. Furthermore, the transition amplitude connecting the degenerate CDW phases with opposite polarizations is much smaller than those connecting the degenerate CP phases. The coherence recovery is achieved with much more difficulty in the CDW phase. 

\section{Concluding Remarks}

We have summarized the history of theories for photoinduced phase transitions, focusing on the development from stochastic to coherent transition dynamics. Both dynamics are characterized by cooperativity. Coherent dynamics are direct consequences of interactions among electrons and/or between electrons and lattice displacements. Photoinduced transition dynamics between different electronic phases of different materials will be studied further in appropriate models for understanding each mechanism and for exploring new possibilities. Current and future issues to be solved will include differences between photoinduced and thermally induced phases, origins of different nonlinear characters, coherence originating from phonons and/or that from electronic excitations, relative dynamics among spins, charge and lattice, relations to quantum phase transitions or quantum natures of phonons, and how to control the dynamics by tuning the laser pulse.

\section*{Acknowledgment}

The authors are grateful to P. Huai, J. Kishine, M. Kuwabara, T. Luty, N. Maeshima, N. Miyashita and H. Zheng for theoretical collaboration, and especially H. Cailleau, S. Koshihara and H. Okamoto among many others for showing their data prior to publication and for enlightening discussions. This work was supported by the NAREGI Nanoscience Project and Grants-in-Aid from the Ministry of Education, Culture, Sports, Science and Technology, Japan.

\end{document}